\documentclass[superscriptaddress,secnumarabic,
amssymb,amsmath,nobibnotes,aps,prd,showpacs,nofootinbib]{revtex4}%
\usepackage{graphicx}
\usepackage{epsf}
\usepackage{bm}
\usepackage{amsmath}
\usepackage{amsfonts}
\usepackage{amssymb}
\usepackage{epstopdf}
\usepackage{natbib}
\usepackage{color}%
\providecommand{\U}[1]{\protect\rule{.1in}{.1in}}
\newcommand{\be}{\begin{equation}}
\newcommand{\ee}{\end{equation}}

\newcommand{\mincir}{\raise
-3.truept\hbox{\rlap{\hbox{$\sim$}}\raise4.truept\hbox{$<$}\ }}
\newcommand{\magcir}{\raise
-3.truept\hbox{\rlap{\hbox{$\sim$}}\raise4.truept\hbox{$>$}\ }}

\newtheorem{remark}{Remark}[section]

\begin{document}



\title{Simple  inflationary quintessential model II: Power law potentials}



\author{Jaume de Haro\footnote{E-mail: jaime.haro@upc.edu}}
\affiliation{Departament de Matem\`atiques, Universitat Polit\`ecnica de Catalunya, Diagonal 647, 08028 Barcelona, Spain} \author{Jaume Amor\'os\footnote{E-mail: jaume.amoros@upc.edu}}
\affiliation{Departament de Matem\`atiques, Universitat Polit\`ecnica de Catalunya, Diagonal 647, 08028 Barcelona, Spain}
\author{Supriya Pan\footnote{E-mail: span@iiserkol.ac.in}}
\affiliation{Department of Physical Sciences, Indian Institute of Science Education and
	Research -- Kolkata, Mohanpur -- 741246, West Bengal, India}


\thispagestyle{empty}

\begin{abstract}

The present work is a sequel of our previous work Phys.\ Rev.\ D {\bf 93}, 084018 (2016) [arXiv:1601.08175 [gr-qc]] which depicted a simple version of an inflationary quintessential model whose inflationary stage was described by a Higgs type potential and the quintessential phase was responsible due to an exponential potential. Additionally, the model predicted a nonsingular universe in past which was geodesically past incomplete. Further, it was also found that the model is in agreement with the Planck 2013 data when running is allowed. But, this model {provides a  theoretical value of the running which is far smaller 
than the central value of the best fit in $(n_s,r,\alpha_s\equiv dn_s/dlnk)$ parameter
space where $n_s$, $r$, $\alpha_s$ respectively denote the spectral index, tensor-to-scalar ratio and the running of the spectral index associated with some inflationary model, and consequently  to analyze the  viability of the model one has to focus in the $2$-dimensional  marginalized confidence level in the allowed domain of the plane $(n_s,r)$ without taking into account the running. Unfortunately, such analysis shows that 
this model does not pass this test.} 
However, in this sequel we propose a family of models runs by a single parameter $\alpha \in [0, 1]$ which proposes another ``inflationary quintessential model'' where the inflation and the quintessence regimes are respectively described by a power law potential and a cosmological constant. The model is also nonsingular although geodesically past incomplete as in the cited model. 
{Moreover,} the present one is found to be more simple in compared to the previous model and it is in excellent agreement with the observational data. 
{In fact, we note that unlike the previous model,
a large number of the models of this family with $\alpha\in [0,\frac{1}{2})$ match with both Planck 2013 and Planck 2015 data without allowing the running}.
 Thus, the properties in the current family of models in compared to its past companion justify its need for a better cosmological model with the successive improvement of the observational data.

\end{abstract}

\vspace{0.5cm}


\maketitle

\section{ Introduction}

In \cite{hap} the authors of the current work
presented a cosmological background in a spatially flat Friedmann - Lema\^{\i}tre - Robertson - Walker (FLRW)
universe  whose dynamics was characterized by
the Raychauduri equation $\dot{H}=F(H)$ where $F (H)$ ($H$ is the Hubble rate of the FLRW universe) was a linear 
function in $H$ before the phase transition and after the phase transition it became quadratic in $H$. In summary, the model realized three fold properties: (i) The linear part prevented the big bang singularity in finite cosmic time although it was geodesically past incomplete, (ii) the phase transition was essential to produce enough particles to reheat the
universe and the universe went through a deflationary period for a sufficient time, and finally (iii) the quadratic part had a fixed point that became responsible for the current acceleration of the universe.
What was worth interesting of that background is that it comes from a quintessential potential, whose inflationary part was a Higgs-style potential and the quintessential part reads an exponential potential. However, the model has some undesired features such as, it provides a reheating temperature in the MeV
regime, although it does not contradict the nucleosynthesis success since this needs a very low temperature, and the worse thing in the model is that, 
since the theoretical value of the running is  far from the corresponding observational mean value obtained by Planck's team, thus, comparing  the theoretical results provided by this model  with Planck 2013  and  Planck 2015 observational data \cite{Ade, Planck} when the running is not allowed, 
one can show that the model has to be disregarded.


However, we found that the existing disparities in \cite{hap} can be defeated in a family of models in this flat FLRW background which has the same feature as in \cite{hap} but in an improved manner. Hence, in this sequel we propose a family of backgrounds whose dynamics before the phase transition is governed by the Raychaudhuri equation $\dot{H}=-k^2 H^{\alpha}$ with $\alpha\in [0,1]$ and $k$ is any real number.
This family provides an inflationary quintessential potential whose inflationary part is basically a power law potential and the quintessential potential is governed by a cosmological constant. Further, the models of this family are nonsingular in nature. In other words, although the models are geodesically past incomplete but they do not encounter with any finite time past singularity, i.e. big bang. Also, the models provide a complete analytic background similar to \cite{hap}.  Further, due to presence of the power law potential,
the models provide a greater reheating temperature in the GeV or TeV regime depending on the value of the parameter $\alpha$.
{Moreover, for some values of this parameter the models match correctly with Planck 2013 \cite{Ade} and Planck 2015 data \cite{Planck}
without allowing the running, 
which indeed is an interesting and notable point in the present family of models.} \newline

The manuscript is organized as follows: In section \ref{model},  we introduce  the family of backgrounds and discuss its properties.  {{}In section \ref{scalar-field}, we establish
that the dynamics governed by the model} could be mimicked by a
single scalar field whose potential is a combination of a {power law potential}, and a {cosmological constant}. Section { 4} is devoted to the study of cosmological
perturbations showing that the
theoretical results provided by our models fit well with current observed data \cite{Ade,Planck}. The reheating process via gravitational particle production  of heavy massive
particles is studied
in section {5}, where we show that our  family of models provide a reheating temperature
in the GeV and TeV regime depending on the value of the parameter $\alpha$.
A detailed calculation  of the number of $e$-fold is performed in section {6}.  In section {7} we compare our new family of models with the
model proposed in our previous work \cite{hap}. Finally in section {8} we have summarized our results.

\vskip 0.3cm

We note that the units used throughout the paper are $\hbar=c=1$.

\section{The model}
\label{model}

We start with the following dynamical equation
\begin{eqnarray}\label{dynamics}
\dot{H}=\left\{\begin{array}{ccc}
(-3H_E^2+\Lambda)\left(\frac{H}{H_E} \right)^{\alpha}& \mbox{for}& H\geq H_E\\
-3H^2+\Lambda& \mbox{for}& H\leq H_E,
\end{array}\right.
\end{eqnarray}
where $H_E$ is a specific value of the Hubble parameter, $\Lambda\ll H_E^2$ is a positive cosmological constant and  $\alpha\in [0,1]$ is the parameter which defines the family of models under consideration.
Now, equation (\ref{dynamics}) can  analytically be solved leading to the following backgrounds:

\begin{enumerate}
	\item For $\alpha=1$ the Hubble parameter is given by
\begin{eqnarray}
 H(t)=\left\{\begin{array}{cc}
      H_Ee^{\frac{(-3H^2_E+\Lambda)t}{H_E}} & t\leq 0\\
      \sqrt{\frac{\Lambda}{3}}\frac{3H_E+\sqrt{3\Lambda}\tanh(\sqrt{3\Lambda} t)}{3H_E\tanh(\sqrt{3\Lambda} t)+\sqrt{3\Lambda}}& t\geq 0,
             \end{array}\right.
\end{eqnarray}
and thus, the scale factor can be solved as
\begin{eqnarray}
 a(t)\cong \left\{\begin{array}{cc}
      a_E e^{\frac{H_E^2}{-3H_E^2+\Lambda}\left[e^{\frac{(-3H_E^2+\Lambda)t}{H_E}}-1\right]} & t\leq 0\\
      a_E \left(\frac{3H_E}{\sqrt{3\Lambda}}\sinh(\sqrt{3\Lambda} t)+\cosh(\sqrt{3\Lambda} t)\right)^{\frac{1}{3}}& t\geq 0.
             \end{array}\right.
\end{eqnarray}

\item For $0\leq \alpha<1$, the Hubble parameter has the following expression
\begin{eqnarray}
 H(t)=\left\{\begin{array}{cc}
      H_E\left((\alpha-1)\left(-3H_E+\frac{\Lambda}{H_E}  \right)t +1\right)^{\frac{1}{1-\alpha}} & t\leq 0\\
      \sqrt{\frac{\Lambda}{3}}\frac{3H_E+\sqrt{3\Lambda}\tanh(\sqrt{3\Lambda} t)}{3H_E\tanh(\sqrt{3\Lambda} t)+\sqrt{3\Lambda}} & t\geq 0,
             \end{array}\right.
\end{eqnarray}
and the corresponding scale factor is
\begin{eqnarray}
 a(t)=\left\{\begin{array}{cc}
      a_E e^{\frac{H_E}{\left(-3H_E+\frac{\Lambda}{H_E}  \right)(2-\alpha)}\left[\left((\alpha-1)
      \left(-3H_E+\frac{\Lambda}{H_E}  \right)t +1\right)^{\frac{2-\alpha}{1-\alpha}} -1\right]} & t\leq 0\\
    a_E \left(\frac{3H_E}{\sqrt{3\Lambda}}\sinh(\sqrt{3\Lambda} t)+\cosh(\sqrt{3\Lambda} t)\right)^{\frac{1}{3}}  & t\geq 0.
             \end{array}\right.
\end{eqnarray}
\end{enumerate}

In all cases ($0\leq \alpha\leq 1$), the family
depicts a nonsingular background in cosmic time
satisfying $H(-\infty)=\infty$, and $H(\infty)=\sqrt{\frac{\Lambda}{3}}$.
Moreover, for $\Lambda\cong 0$, one can have the following approximate forms of the Hubble parameter and the scale factor.
\begin{enumerate}
	\item For $\alpha=1$:

\begin{eqnarray}
 H(t)\cong\left\{\begin{array}{cc}
      H_Ee^{-3H_Et} & t\leq 0\\
      \frac{H_E}{3H_Et+1}& t\gtrsim 0,
             \end{array}\right.
\end{eqnarray}
and
\begin{eqnarray}
 a(t)\cong \left\{\begin{array}{cc}
      a_E e^{-\frac{1}{3}[e^{-3H_Et}-1]} & t\leq 0\\
      a_E (3H_Et+1)^{\frac{1}{3}}& t\gtrsim 0.
             \end{array}\right.
\end{eqnarray}
\item For $0\leq \alpha<1$:

\begin{eqnarray}
 H(t)=\left\{\begin{array}{cc}
     H_E\left(3(\alpha-1)H_Et +1\right)^{\frac{1}{1-\alpha}}   & t\leq 0\\
    \frac{H_E}{3H_Et+1}& t\gtrsim 0,
             \end{array}\right.
\end{eqnarray}
and
\begin{eqnarray}
 a(t)=\left\{\begin{array}{cc}
      a_E e^{-\frac{1}{3(2-\alpha)}[\left(3(\alpha-1)H_Et +1\right)^{\frac{2-\alpha}{1-\alpha}} -1]} & t\leq 0\\
      a_E (3H_Et+1)^{\frac{1}{3}}& t\gtrsim 0.
             \end{array}\right.
\end{eqnarray}
\end{enumerate}

On the other hand,
the effective Equation of State (EoS) parameter, namely $w_{eff}$, which is defined as $w_{eff}=-1-\frac{2\dot{H}}{3H^2}$, for our  family of models is given by
\begin{eqnarray}
 w_{eff}=\left\{\begin{array}{cc}
      -1+2\left(1-\frac{\Lambda}{3H_E^2} \right)  \left(\frac{H}{H_E} \right)^{\alpha-2} & H\geq H_E\\
      1- \frac{2\Lambda}{3H^2}&  H\leq H_E,
             \end{array}\right.
\end{eqnarray}
which shows that for $H\gg H_E$ one has
$w_{eff}(H)\cong -1$ (early quasi de Sitter period). When $H\cong H_E$, the EoS parameter satisfies
$w_{eff}(H)\cong 1$ (kination or deflationary period \cite{spokoiny,joyce}), and finally, for $H\cong  \sqrt{\frac{\Lambda}{3}}$ one also has  $w_{eff}(H)\cong -1$ (late quasi
de Sitter period).
Moreover, when one considers the approximation $\Lambda=0$, the Equation of State becomes
\begin{eqnarray}
 P=\left\{\begin{array}{cc}
      -\rho+2\left(\frac{\rho}{\rho_E} \right)^{\frac{\alpha-2}{2}}\rho & \rho\geq \rho_E\\
      \rho&  \rho\leq \rho_E .
             \end{array}\right.
\end{eqnarray}
where {$\rho$, $P$ are respectively the energy density and the pressure of the cosmic fluid} and $\rho_E$ is the energy density of the universe at $H= H_E$.  In particular,

\begin{enumerate}\item
For $\alpha=0$, the equation of state becomes
\begin{eqnarray}\label{eos1}
 P=\left\{\begin{array}{cc}
      -\rho+2{\rho_E}  & \rho\geq \rho_E\\
      \rho&  \rho\leq \rho_E .
             \end{array}\right.
\end{eqnarray}

\item
For $\alpha=1$, the equation of state takes the form
\begin{eqnarray}\label{eos2}
 P=\left\{\begin{array}{cc}
      -\rho+2\sqrt{\rho\rho_E } & \rho\geq \rho_E\\
      \rho&  \rho\leq \rho_E .
             \end{array}\right.
\end{eqnarray}
\end{enumerate}

\section{The scalar field}
\label{scalar-field}

{It is evident from equations (\ref{eos1}) and (\ref{eos2}) that
at early times, our family of backgrounds  satisfies $P(\rho)\cong -\rho$, that means our universe was quasi de Sitter in nature. Since the dynamics of the early accelerating phase is well realized via a scalar field prescription, hence it is very natural to ask whether an equivalence between the family and the scalar field dynamics exists or not. If such an equivalence exists then we need to confirm their viability with the observational data, that means essentially we aim to check whether the family of models} could lead to a power spectrum of cosmological perturbations that fit well with the current observational data \cite{Ade, Planck}.
To do so, in the flat FLRW universe, if we represent the energy density and the pressure by the notations
$\rho_{\varphi}$, $p_{\varphi}$, respectively, then they assume the following simplest forms
{
\begin{align}\label{sf1}
\rho_{\varphi} & = \frac{\dot{\varphi}^2}{2}+ V (\varphi),~~~p_{\varphi}= \frac{\dot{\varphi}^2}{2}- V (\varphi)
\end{align}
}

Now, using Eq. (\ref{sf1}) and the Raychaudhuri equation {{} $\dot{H}=-\frac{\dot{\varphi}^2}{2M_{pl}^2} $} (where $M_{pl}^2= (8 \pi G)^{-1}$, is the reduced Planck's mass), we find
\begin{eqnarray}
\varphi=M_{pl}\int\sqrt{-2\dot{H}}\,\,dt=-M_{pl}\int\sqrt{-\,\,\frac{2}{\dot{H}}}\,\,dH.
\end{eqnarray}
Now, in our case the scalar field is solved as

\begin{eqnarray}
 \varphi=\left\{\begin{array}{cc}
    \varphi_E\left(\frac{H}{H_E}\right)^{\frac{2-\alpha}{2}}
    & H\geq H_E \\
    -\sqrt{\frac{2}{3}}M_{pl}\ln\left(\frac{\sqrt{H^2-\frac{\Lambda}{3}} +H }{  \sqrt{H_E^2-\frac{\Lambda}{3}} +H_E        }\right) +\varphi_E& H \lesssim  H_E,
                \end{array}\right.
\end{eqnarray}
where  $\varphi_E\equiv -\frac{2\sqrt{2}}{\sqrt{3}(2-\alpha)}\frac{H_E}{\sqrt{H_E^2-\frac{\Lambda}{3}}}M_{pl}
\cong -\frac{2\sqrt{2}}{\sqrt{3}(2-\alpha)}M_{pl}$.

Conversely, one can express the Hubble rate in terms of the field as
\begin{eqnarray}
 H=\left\{\begin{array}{cc}
    H_E\left(\frac{\varphi}{\varphi_E}\right)^{\frac{2}{2-\alpha}}& \varphi\leq \varphi_E \\
     \frac{(\sqrt{H_E^2-\frac{\Lambda}{3}} +H_E    )^2e^{-\sqrt{\frac{3}{2}}(\varphi-\varphi_E)}+\frac{\Lambda}{3}e^{\sqrt{\frac{3}{2}}(\varphi-\varphi_E)}
     }{2(\sqrt{H_E^2-\frac{\Lambda}{3}} +H_E)    }& \varphi \geq \varphi_E.
                \end{array}\right.
\end{eqnarray}

The potential is given by $V(H)=3H^2M_{pl}^2+\dot{H}M_{pl}^2\Longrightarrow V(\varphi)=3H^2(\varphi)M_{pl}^2+\dot{H}(\varphi)M_{pl}^2.$ Then, for our family one has
\begin{eqnarray}
 V(H)=\left\{\begin{array}{cc}
           3H^{\alpha}\left(H^{2-\alpha}-\frac{H_E^{2}-\frac{\Lambda}{3}}{H_E^{\alpha}}\right)M_{pl}^2& H\geq H_E\\
           \Lambda M_{pl}^2& H\leq H_E.
          \end{array} \right.
\end{eqnarray}

That is,

\begin{eqnarray}
 V(\varphi)=\left\{\begin{array}{cc}
           3\left(\frac{H_E M_{pl}}{\varphi_E} \right)^2\left( \frac{\varphi}{\varphi_E}\right)^{\frac{2\alpha}{2-\alpha}}
           \left[\varphi^2-\varphi_E^2\left(1- \frac{\Lambda}{3H_E^2} \right)\right]& \varphi\leq \varphi_E\\
          \Lambda M_{pl}^2 & \varphi\geq \varphi_E.
          \end{array} \right.
\end{eqnarray}

Note that, for $\alpha=0$, the potential is quadratic, for $\alpha=\frac{2}{3}$, it is cubic and for $\alpha=1$, it is quartic.

\section{Cosmological perturbations}
Now, to study the cosmological perturbations, one needs to introduce the slow roll parameters \cite{btw}
\begin{eqnarray}\label{slowroll}
 \epsilon=-\frac{\dot{H}}{H^2}, \quad \eta=2\epsilon-\frac{\dot{\epsilon}}{2H\epsilon},
\end{eqnarray}
which allow us to calculate the associated inflationary parameters, such as, the spectral index ($n_s$), its running ($\alpha_s$), and the ratio of tensor to scalar perturbations ($r$) defined below
\begin{eqnarray}
 n_s-1=-6\epsilon_*+2\eta_*, \quad \alpha_s=\frac{H\dot{n}_s}{H^2 +\dot{H}},\quad
 r=16\epsilon_*
\end{eqnarray}
where the star ($*$) means that the quantities are evaluated when the pivot scale crosses the Hubble radius.
{{}
Now, for our family of models, the above inflationary parameters assume the following values
\begin{eqnarray}
n_s-1=(\alpha-4)\epsilon_*, \quad \alpha_s=\frac{(\alpha-4)(2-\alpha)\epsilon^2_*}{1-\epsilon_*},\quad
 r=16\epsilon_*,
\end{eqnarray}
where $\epsilon_*=3\left(\frac{H_E}{H_*} \right)^{2-\alpha}$.  Now, let us remark the following:

\begin{remark}
 For potentials of the form $V(\varphi)=\lambda \varphi^{\frac{4}{2-\alpha}}$, and using that

$$\epsilon \cong \frac{M_{pl}^2}{2}\left( \frac{V_{\varphi}}{V} \right)^2, \qquad
\eta \cong {M_{pl}^2}\frac{V_{\varphi\varphi}}{V} $$
one also obtains that
$n_s-1\cong (\alpha-4)\epsilon_*$,
which means that our family of potentials, during the inflationary regime, are like power law potentials
\end{remark}

The number of $e$-folds is given by
\begin{eqnarray}
N=\int_{t_*}^{t_{end}} H dt=-\int_{H_{end}}^{H_*} \frac{H}{\dot{H}}dH=\frac{1}{2-\alpha}\left(\frac{1}{\epsilon_*}-1 \right).
\end{eqnarray}
Then, in terms of the number of $e$-folds one has
\begin{eqnarray}
 n_s-1=\frac{\alpha-4}{1+(2-\alpha)N}, \quad
 r= \frac{16}{1+(2-\alpha)N}, \quad
 \alpha_s=\frac{\alpha-4}{N(1+(2-\alpha)N}.
\end{eqnarray}

{From this last formula and  due to the large value of the number of $e$-folds, one can see that the running is of the same order as
$(n_s-1)^2$. Then, it is clear that its theoretical value is far smaller than the central value of the best fit obtained by Planck's team (see for instance Table 5 of \cite{Ade}),
and as a consequence to  check the viability of our models we have to consider the 2-dimensional marginalized confidence level in the plane $(n_s,r)$  without the presence of running provided by
 Planck's team. Moreover, it is important to realize that in quintessential inflation, the number
 of $e$-folds is greater than the $e$-folds
 for inflationary potentials with a deep well \cite{ll}. For this reason, here we have drawn the curves from $N=65$ to $N=75$. Taking  into account these considerations, we have showed in
figure $1$  that the  models allowed by Planck 2015 data at $2\sigma$ C.L. must satisfy $\alpha\in[0,\frac{1}{2})$.
}

\begin{figure}
\includegraphics[height=0.55\textwidth,angle=0]{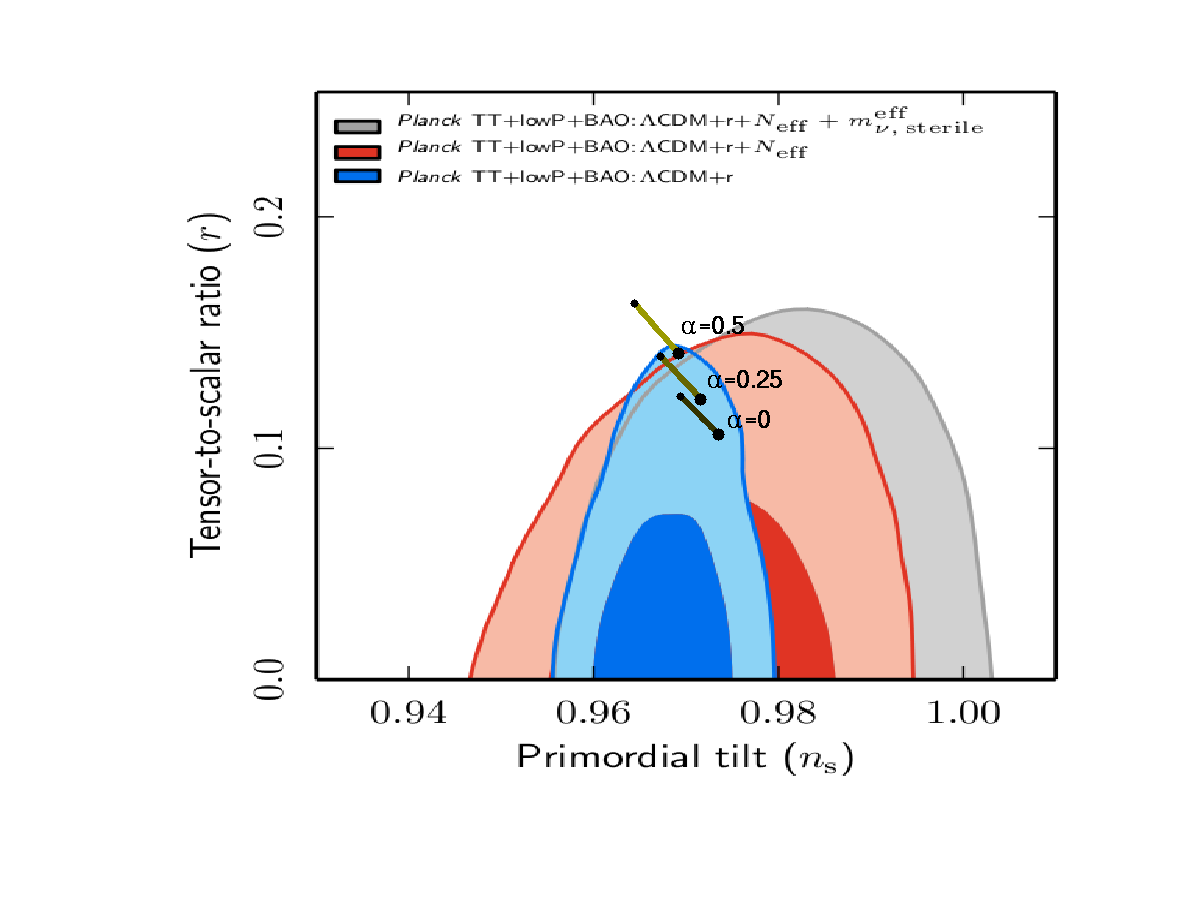}
\caption{Marginalized joint confidence contours for $(n_{\mathrm s} \,, r)$,
at the 68\,\% and 95\,\% CL,  without the presence of running of the spectral indices. For the values $\alpha=0, \frac{1}{4}$ and $\frac{1}{2}$ we have drawn the curves from $65$
(small circle) to
$75$ (big circle) $e$-fols.
({Figure courtesy of the Planck2015 Collaboration}).
}
\label{fig:nsrr}
\end{figure}

\vspace{0.5cm}
Finally,
to determinate the value of $H_E$, one has to take into account the theoretical \cite{btw} and the observational \cite{bld} value of the power spectrum
\begin{eqnarray}\label{power1}
 {\mathcal P}\cong \frac{H^2}{8\pi^2\epsilon_* M_{pl}^2}\cong 2\times 10^{-9}.
\end{eqnarray}

Using  $H_*=\frac{H_E}{\left(\frac{\epsilon_*}{3} \right)^{\frac{1}{2-\alpha}}}$ and  $\epsilon_*=\frac{1-n_s}{4-\alpha}$, one obtains
\begin{eqnarray}
 H_E \sim 7\times 10^{-4}
\left(\frac{1-n_s}{3(4-\alpha)} \right)^{\frac{4-\alpha}{2(2-\alpha)}} M_{pl}.
\end{eqnarray}

Taking, as usual, $n_s\cong 0.96$ one has the value of $H_E$ for each value of the parameter $\alpha$.
In particular:
 For $\alpha=0$, one has $H_E\sim  2\times 10^{-6} M_{pl}\sim 5\times 10^{12}$ GeV,
 and for $\alpha=1$, one has $H_E\sim  10^{-7} M_{pl}\sim  2\times 10^{11}$ GeV.

}

\section{The reheating process}
{
We devote this section on the production of heavy massive particles ($m\gg H_E$) which are conformally coupled to gravity due to a phase transition to a deflationary regime \cite{hap,he}. In this case, since the second derivative of the Hubble parameter is discontinuous,
during the adiabatic regimes,
we will use the first order WBK solution to define the approximate vacuum modes \cite{Haro}
\begin{eqnarray}
\chi_{1,k}^{WKB}(\tau)\equiv
\sqrt{\frac{1}{2W_{1,k}(\tau)}}e^{-{i}\int^{\tau}W_{1,k}(\eta)d\eta},
\end{eqnarray}
where $W_{1,k}$ can be calculated as
\begin{eqnarray}
W_{1,k}=
\omega_k-\frac{1}{4}\frac{\omega''_{k}}{\omega^2_{k}}+\frac{3}{8}\frac{(\omega'_{k})^2}{\omega^3_{k}} .
\end{eqnarray}

Now, before the phase transition the vacuum is depicted approximately by $\chi_{1,k}^{WKB}(\tau)$, but after the phase transition this mode becomes a mixture of
positive and negative frequencies of the form
$\alpha_k \chi_{1,k}^{WKB}(\tau)+\beta_k (\chi_{1,k}^{WKB})^*(\tau)$.

The $\beta_k$-Bogoliubov coefficient could be obtained, as usual, matching both expressions at the transition time $\tau_E$, obtaining
{\begin{eqnarray*}
\beta_k=\frac{{\mathcal W}[\chi_{1,k}^{WKB}(\tau_E^-),\chi_{1,k}^{WKB}(\tau_E^+)]}
{{\mathcal W}[(\chi_{1,k}^{WKB})^*(\tau_E^+),\chi_{1,k}^{WKB}(\tau_E^+)]},~~ \mbox{where}~{\mathcal W}~\mbox{is the Wronskian}.
\end{eqnarray*}}

The square modulus of the {}{$\beta_k$-Bogoliubov} coefficient will be given by
{\begin{eqnarray*}
 |\beta_k|^2\cong \frac{m^4a_E^{10}\left(\ddot{H}_E^+-\ddot{H}_E^-\right)^2}{256(k^2+m^2a^2_E)^5}=
 \frac{81 (2-\alpha)^2m^4a_E^{10}H_E^6}{256(k^2+m^2a^2_E)^5}.
\end{eqnarray*}}

The number and energy density are given by
\begin{eqnarray}
n_{\chi}\equiv \frac{1}{2\pi^2 a^3}\int_0^{\infty}k^2|\beta_k|^2dk,\quad
\rho_{\chi}\equiv \frac{1}{2\pi^2 a^4}\int_0^{\infty}k^2\omega_k|\beta_k|^2dk.
\end{eqnarray}

Then for our family one has
\begin{eqnarray}
 n_{\chi}\sim 3\times 10^{-3}(2-\alpha)^2\frac{H_E^6}{m^3}\left(\frac{a_E}{a} \right)^3, \quad \rho_{\chi}\sim mn_{\chi}.
\end{eqnarray}

{We notice that at the beginning of reheating, the particles  are far from being in thermal equilibrium,} 
and at first their energy density scales
as $a^{-3}$, eventually they will decay into lighter
particles, which will interact through multiple scattering. At the end of this process, the universe becomes filled with a
relativistic plasma in thermal equilibrium whose energy density decays as $a^{-4}$. Now, since the energy density of the background decays as $a^{-4}$ (i.e. deflationary regime), eventually the
energy density of the relativistic plasma will dominate and the universe will become reheated.

Here, as in \cite{pv, 37}, we consider the thermalization process,  where the cross section for $2\rightarrow 3$ scattering with  gauge bosons exchange whose typical energy is $\rho_{\chi}^{\frac{1}{4}}(0) $,
is given by
$\sigma={\beta^3}\rho_{\chi}^{-\frac{1}{2}}(0)$, with $\beta^2\sim 10^{-3}$. The thermalization rate is
\begin{eqnarray*}
 \Gamma=\sigma n_{\chi}(0)
 \sim 5\times 10^{-2}(2-\alpha)\beta^3\left(\frac{H_E}{m}\right)^2H_E.
\end{eqnarray*}
  Thermal equilibrium is reached when {}{$\Gamma\sim H(t_{eq})\cong H_E\left(\frac{a_E}{a_{eq}}\right)^3$},  which leads to the relation {}
  {$\frac{a_E}{a_{eq}}\sim 4\times 10^{-1}(2-\alpha)^{1/3}\beta \left(\frac{H_E}{m}\right)^{2/3}$}. Then, at the equilibrium one has
\begin{eqnarray}
 \rho_{\chi}(t_{eq})\sim 10^{-4}(2-\alpha)^3\beta^3\left(\frac{H_E}{m}\right)^{4}H_E^4,\qquad
 \rho(t_{eq})\sim 7\times 10^{-3}(2-\alpha)^2\beta^6\left(\frac{H_E}{m}\right)^{4}H_E^2 M_{pl}^2.
\end{eqnarray}

After this thermalization, the relativistic plasma and the background  evolve as {}
{\begin{eqnarray}\rho_{\chi}(t)=\rho_{\chi}(t_{eq})\left(\frac{a_{eq}}{a} \right)^4,
\quad \rho(t)=\rho(t_{eq})\left(\frac{a_{eq}}{a} \right)^6,\end{eqnarray}}
and the reheating is obtained when both energy densities are
of the same order,  which happens when {}{$\frac{a_{eq}}{a_R}\sim \sqrt{\frac{\rho_{\chi}(t_{eq})}{\rho(t_{eq})}}$}, and thus,
one obtains a  reheating temperature of the order
\begin{eqnarray*}
 T_R\sim \rho_{\chi}^{\frac{1}{4}}(t_{eq})\sqrt{\frac{\rho_{\chi}(t_{eq})}{\rho(t_{eq})}} \sim 
 10^{-1}\left(\frac{H_E}{M_{pl}} \right)^2\left(\frac{H_E}{m}\right)M_{pl}.
\end{eqnarray*}

Since,  $H_E\ll m$, if we consider masses of the order $10^2 H_E$ one has
$$T_R\sim 10^{-3}\left(\frac{H_E}{M_{pl}} \right)^2M_{pl}\sim 5\times 10^{-10} \left(\frac{1-n_s}{3(4-\alpha)} \right)^{\frac{4-\alpha}{2-\alpha}} M_{pl}. $$

As a particular cases we consider
\begin{enumerate}
 \item The  quadratic potential corresponding to $\alpha=0$, leads to the reheating temperature $T_R\sim 5\times 10^{-15}M_{pl}\sim  10^4$ GeV.
 \item  The cubic potential corresponding to $\alpha=\frac{2}{3}$, leads to the reheating temperature $T_R\sim 5\times10^{-16}M_{pl}\sim 10^3$ GeV.
\item  The quartic potential corresponding to $\alpha=1$, leads to the reheating temperature $T_R\sim  4\times10^{-17}M_{pl}\sim 10^2$ GeV.
\end{enumerate}

}

\vspace{1cm}

Finally, to end this section,  we study the evolution after reheating. Since after the phase transition the potential is constant one will have
\begin{eqnarray}
 \ddot{\varphi}+3H\dot{\varphi}=0\Longleftrightarrow \dot{\varphi}(t)=\dot{\varphi}(t_R)e^{-3\int_{t_R}^tH(s)ds},
\end{eqnarray}
where $t_R$ is the reheating time.

On the other hand, during the radiation and the matter dominated phases,  one will have
\begin{eqnarray}
 H(t)=\frac{H_R}{1+2(t-t_R)H_R}, \quad \mbox{and} \quad H(t)=\frac{2H_M}{2+3(t-t_M)H_M},
\end{eqnarray}
where the subindices $R$, $M$ respectively denote the Hubble rate when radiation and matter domination will start to dominate. Then if we denote by $t_{\Lambda}$ the time when the cosmological constant starts to
dominate on, one will get
\begin{eqnarray}
 \dot{\varphi}(t_{\Lambda})=\frac{\dot{\varphi}(t_R)}{(1+2(t_M-t_R)H_R)^{\frac{3}{2}}(2+3(t_{\Lambda}-t_M)H_M)^2}.
\end{eqnarray}

Since nowadays the universe is accelerating one can take $\Lambda\sim H_0^2$, where $H_0$ is the current value of the Hubble parameter, and thus, one arrives at
\begin{eqnarray}
 \dot{\varphi}^2(t_{\Lambda})\sim \dot{\varphi}^2(t_R)\frac{H_MH_0^2}{H_R^3}.
\end{eqnarray}

As a consequence, since at the beginning of the radiation domination, all the energy density is kinetic,  the ratio between the kinetic and potential energy density ($\mathcal{R}$) when the cosmological constant starts to dominate satisfies
\begin{eqnarray}
\mathcal{R}\cong \frac{\dot{\varphi}^2(t_{\Lambda})/2}{\Lambda M_{pl}^2}\sim \frac{H_M}{H_R}.
\end{eqnarray}

Now using that the value of the Hubble parameter at the beginning of the matter domination is of the order $H_M\sim 10^{-54}M_{pl}$ (see \cite{he}) and for
our models $H_R$ belongs between $10^{-30}M_{pl}$ and $10^{-34} M_{pl}$, one can calculate that
\begin{eqnarray}
 \mathcal{R}\leq 10^{-20},
\end{eqnarray}
which means that the kinetic part of the energy density is sub-dominant, and thus, in our model,  it is the cosmological cosmological constant which drives the current evolution of the universe.

\section{Calculation of the number of $e$-folds}
{
We start with the main formula \cite{ll}
\begin{eqnarray}
\frac{k_*}{a_0H_0}
=e^{-N_*}\frac{H_*}{H_0}\frac{a_{end}}{a_E}\frac{a_E}{a_R}\frac{a_R}{a_M}\frac{a_M}{a_0}
=e^{-N_*}\frac{H_*}{H_0}\frac{a_{end}}{a_E}
\frac{\rho_R^{-1/12}\rho_M^{1/4}}{\rho^{1/6}_E}\frac{a_M}{a_0},
\end{eqnarray}
where ``end'', $R$ and $M$ respectively symbolize the end of inflation,  the beginning of radiation era,  and the beginning of the matter domination era. Further, the subindex `$0$' at any quantity means its value at current time. Here we have used the relations
\begin{eqnarray}
\left(\frac{a_{E}}{a_R}\right)^{6}=\frac{\rho_R}{\rho_{E}}, \quad
\left(\frac{a_R}{a_M}\right)^4=\frac{\rho_M}{\rho_R}.
\end{eqnarray}

Taking the pivot scale as $k_*=0.05$ $\mbox{Mpc}^{-1}$, and since the current horizon scale is
 $a_0H_0\cong 2\times 10^{-4}$ $\mbox{Mpc}^{-1}$, one obtains

 \begin{eqnarray}
N_*=-5.52+\ln\left(\frac{H_*}{H_0} \right)+\ln\left(\frac{a_{end}}{a_E} \right)
+\frac{1}{4}\ln\left(\frac{\rho_M}{\rho_R} \right) +\frac{1}{6}\ln\left(\frac{\rho_R}{\rho_E} \right)
+\ln\left(\frac{a_M}{a_0} \right).
\end{eqnarray}

 Since after reheating, the process becomes
adiabatic, i.e. $T_0=\frac{a_M}{a_0}T_M$, hence  using the relations $\rho_M\cong \frac{\pi^2}{15}g_M T_M^4$
 and $\rho_R\cong \frac{\pi^2}{30}g_R T_R^4$ {(where $g_i$'s, $i= R, M$ are the relativistic degrees of freedom\footnote{Specifically $g_M$ stands for the number of relativistic degrees of freedom at matter radiation equality and $g_R$ is the number of relativistic degrees of freedom at the end of reheating.} \cite{rg}.)},  one arrives at
 \begin{eqnarray}
N_*=-5.52+\ln\left(\frac{H_*}{H_0} \right)+\ln\left(\frac{a_{end}}{a_E} \right)
+\frac{1}{4}\ln\left(\frac{2g_M}{g_R} \right) +\frac{1}{6}\ln\left(\frac{\rho_R}{\rho_E} \right)
+\ln\left(\frac{T_0}{T_R} \right).
\end{eqnarray}

Now, taking into account that $H_0\sim 6\times 10^{-61} M_{pl}$ and ${\mathcal P}=\frac{H_*^2}{8\pi^2\epsilon_*M_{pl}^2}\sim 2\times 10^{-9}$ one obtains
 \begin{eqnarray}
  \ln\left(\frac{H_*}{H_0} \right)=131.38+\frac{1}{2}\ln\left(\frac{1-n_s}{3(4-\alpha)} \right).
 \end{eqnarray}

 Now using the current temperature of the universe $T_0\cong 2.73$ K $\cong 2\times 10^{-13}$ GeV and $g_M=3.36$ \cite{rg} one has
 \begin{eqnarray}
 \frac{1}{4}\ln\left(\frac{2g_M}{g_R} \right)+\ln\left(\frac{T_0}{T_R} \right)=-28.76-\ln\left(\frac{g_R^{\frac{1}{4}}T_R}{\mbox{GeV}} \right)
 \end{eqnarray}
 From the value of the Hubble parameter at the transition time, one will obtain
 \begin{eqnarray}
 \frac{1}{6}\ln\left(\frac{\rho_R}{\rho_E} \right)=-26.16- \frac{4-\alpha}{6(2-\alpha)}\ln\left(\frac{1-n_s}{3(4-\alpha)} \right)
 +\frac{2}{3}\ln\left(\frac{g_R^{\frac{1}{4}}T_R}{\mbox{GeV}} \right).
 \end{eqnarray}

Collecting all the terms one obtains
 \begin{eqnarray}
N_*=70.94+\ln\left(\frac{a_{end}}{a_E} \right)
+\frac{1-\alpha}{3(2-\alpha)}\ln\left(\frac{1-n_s}{3(4-\alpha)} \right)-\frac{1}{3}\ln\left(\frac{g_R^{\frac{1}{4}}T_R}{\mbox{GeV}} \right)
\end{eqnarray}

On the other hand, a simple calculation leads to
\begin{eqnarray}
\ln\left(\frac{a_{end}}{a_E} \right)=\int_{H_E}^{H_{end}}\frac{H}{\dot{H}}dH=-\frac{2}{3(2-\alpha)}.
\end{eqnarray}

 \begin{eqnarray}
N_*=70.94-\frac{1}{3(2-\alpha)}\left[2-
(1-\alpha)\ln\left(\frac{1-n_s}{3(4-\alpha)} \right)\right]-\frac{1}{3}\ln\left(\frac{g_R^{\frac{1}{4}}T_R}{\mbox{GeV}} \right)
\end{eqnarray}

Finally, since for our models
$T_R\sim  5\times 10^{-10} \left(\frac{1-n_s}{3(4-\alpha)} \right)^{\frac{4-\alpha}{2-\alpha}} M_{pl}   \sim 10^{8}\left(\frac{1-n_s}{3(4-\alpha)} \right)^{\frac{4-\alpha}{2-\alpha}}$ GeV, hence
using the fact that  $g_R=107$ for $T_R\geq 175$ GeV \cite{rg},
one gets
 \begin{eqnarray}
N_*=64.41-\frac{1}{3(2-\alpha)}\left[2+3
\ln\left(\frac{1-n_s}{3(4-\alpha)} \right)\right].
\end{eqnarray}

Taking as usual $n_s\cong 0.96$ we observe the following:
 \begin{enumerate}
  \item  For the quadratic potential ($\alpha=0$), the number of $e$-folds is $N_*=67$.
  \item   For the cubic potential ($\alpha=\frac{2}{3}$), the number of $e$-folds is  $N_*=68$.
  \item  For the quartic potential ($\alpha=1$), the number of $e$-folds is $N_*=69$.
 \end{enumerate}

}

\section{Comparison with the previous model}
In section $7$ of our previous work \cite{hap}, we  introduced the following dynamical system:
\begin{eqnarray}\label{background1}
 \dot{H}=\left\{\begin{array}{ccc}
                 {-3H_e^2}(2H-H_e)& \mbox{for} & H>H_E\\
                 -3H^2+{\Lambda} & \mbox{for} & H\leq H_E,
                \end{array}\right.
\end{eqnarray}
where $H_e$ is the model parameter and the phase transition occurs at $H_E=H_e+\sqrt{\frac{\Lambda}{3}}$. We note that this background originally comes from the following quintessential Higgs-style potential
\begin{eqnarray}
 V(\varphi)=\left\{\begin{array}{ccc}
                 \frac{27H_e^2M_{pl}^2}{16}\left(\frac{\varphi^2}{M_{pl}^2}-\frac{2}{3} \right)^2& \mbox{for} & \varphi<\varphi_E\\
                 {\Lambda}M_{pl}^2 & \mbox{for} & \varphi\geq \varphi_E,
                \end{array}\right.
\end{eqnarray}
in which  $\varphi_E=-M_{pl}\sqrt{\frac{2}{3}}\sqrt{1+\frac{2}{H_e}\sqrt{\frac{\Lambda}{3}}}\cong -M_{pl}\sqrt{\frac{2}{3}}.$

It has been shown in \cite{hap, he} that due to the gravitational production of heavy massive particles, the reheating temperature
belongs in the MeV regime. This is due to the fact that the second derivative of the Hubble parameter is continuous and the third
one is discontinuous at the transition time. However, for the family of models described by the sole parameter $\alpha \in [0, 1]$, the second derivative of the Hubble parameter is discontinuous at the
transition phase what leads to a reheating temperature in the GeV  or TeV regime.

{Moreover,  during the inflationary period, the  Higgs-style potential  has the same behavior as a quartic one (i.e. the model with $\alpha = 1$), which means that 
it does not match with Planck 2015 data (see figure $1$).}
On the contrary, for our new family of models, if one takes $\alpha\in [0,\frac{1}{2})$, the corresponding models match at $2\sigma$ C.L.  with Planck 2013 and Planck 2015 without allowing the  running. {It is one of the main results of the present work which proves the potentiality of the current family of models in compared to our earlier work \cite{hap}.}

\section{Summary and Discussion}
\label{discuss}

The current work offers a sequel of our previous work \cite{hap} with a significant improvement in compared to its mathematical simplicity, and in agreement with the very latest observational data.
Let us demonstrate the main improvements by comparing the previous model \cite{hap} with the current one and finally the need of this sequel.\newline

In \cite{hap} a simple unified cosmological model was proposed having (i) an inflationary period described by a Higgs type potential, (ii) a sudden phase transition from the inflationary phase to the deflationary phase, which results in the production of massive particles, hence the universe begins to reheat, after that, it successively enters into the radiation and matter dominated eras, and finally,  (iii) a quintessential stage explained by an exponential potential. Additionally, although the model was geodesically past incomplete, but it did not encounter any big bang singularity in the finite cosmic time. So, essentially, we realized a singularity free cosmological model unifying
the early inflationary epoch with the current cosmic acceleration by only a single scalar field whose potential is a combination of Higgs potential and an exponential one.
Also, the model provided an complete analytic background which thus was helpful to calculate the associated cosmological parameters.
We found that the model {only agrees with the Planck 2013 data \cite{Ade} in presence of running, 
however, as we have already discussed, 
the model does not match with the Planck's observational data without the presence of running}.\newline



But in the present work we provide an improved version of \cite{hap} which is potential and worthy for further discussions. Here we propose a family of new cosmological models described by a sole parameter $\alpha \in [0, 1]$ which provides a complete picture of our universe via a single scalar field as in \cite{hap}. The current family of models have (i) an inflationary phase described by a power law potential, (ii) a sudden phase transition from inflationary regime to the deflationary regime, hence beginning of reheating, consequently, successive radiation, matter dominated eras, and finally (iii) the models enter into the current accelerating phase responsible by the cosmological constant. In addition to that, the family of models are nonsingular in nature, i.e. they do not predict any finite cosmic time big bang singularity, but are geodesically past incomplete. That means similar to \cite{hap} the present family of models also unifies the inflationary epoch with the current accelerating phase by a combination of a power law potential and a cosmological constant. In particular, the inflationary power law potentials are recognized by the models with $\alpha = 0$, $\frac{2}{3}$, $1$, as quadratic, cubic and quartic potentials in which for the observable modes, the universe inflates respectively for a number of 67, 68, and 69 $e$-folds.
Another interesting point in the family of models is that the reheating temperature could reach the GeV or TeV regime  depending on the value of the sole parameter ``$\alpha$'' unlike in the previous model \cite{hap} where the reheating temperature belongs in the MeV regime. {Finally,  we found that  a large number of models having $\alpha \in [0, \frac{1}{2})$ match both Planck 2013 and Planck 2015 data at 2$\sigma$ C.L.  without the need of running, which does not happen with the model presented in \cite{hap}.} \newline

Summarizing, the current family of models describes a simple nonsingular inflationary quintessential analytic cosmological models, providing a greater reheating temperature in the GeV/TeV regime, and a large number of models belonging to this family are in excellent agreement with the Planck 2015 data without allowing the running, and hence it reports a significant improvement of \cite{hap}.

\section*{Acknowledgments}
{The authors would like to thank the anonymous referee  for his/her valuable suggestions that have been very useful to improve the paper.}
This investigation has been supported in part by MINECO (Spain), projects MTM2014-52402-C3-1-P and MTM2012-38122-C03-01. Research of SP has been funded by the National Post-Doctoral Fellowship (File No: PDF/2015/000640) provided by the Science and Engineering Research Board (SERB), Govt. of India.

\end{document}